\documentclass[twocolumn, a4paper]{article}
\usepackage[utf8]{inputenc}
\usepackage{amssymb,amsmath}
\usepackage{algpseudocode,algorithm,algorithmicx}
\usepackage{subfigure}
\usepackage{xcolor}
\usepackage{comment}
\usepackage{tikz,authblk}
\usepackage[margin=2cm]{geometry}
\usetikzlibrary{matrix,calc}

\newcommand{\proglang}[1]{\texttt{#1}}
\newcommand{\code}[1]{\texttt{#1}}
\newcommand{\fct}[1]{\code{#1()}}

\newcommand{\Real}{\mathbb{R}}

\newcommand{\ie}{\textit{i.e.}}
\newcommand{\eg}{\textit{e.g.}}

\newcommand{\midx}[1]{\mathbf{#1}}
\newcommand{\mat}[1]{\mathbf{#1}}

\newcommand{\tensor}[1]{\mathcal{#1}}

\date{\today}

\title{Algorithm for an arbitrary-order cumulant tensor calculation in a 
sliding window of data streams}

\begin{document}
	\author[1]{Krzysztof Domino}
	\author[1]{Piotr Gawron\thanks{gawron@iitis.pl}}
	
	%\correspondingauthor{Piotr Gawron}
	
	\affil[1]{Institute of Theoretical and Applied Informatics,
		Polish Academy of Sciences,\protect\\ 
		Ba{\l}tycka 5, 44-100 Gliwice, Poland}
	
\maketitle

\begin{abstract}
    High order cumulant tensors carry information about statistics of
    non-normally distributed multivariate data. In this work we present a~new
    efficient algorithm for calculation of cumulants of arbitrary order in a
    sliding window for data streams. We showed that this algorithms enables 
    speedups of cumulants updates compared to current algorithms. This 
    algorithm can be used for processing on-line high-frequency
    multivariate data and can find applications in, e.g., on-line
    signal filtering and classification of data streams.
    To present an application of this algorithm, we propose an
    estimator of non-Gaussianity of a data stream based on the norms
    of high-order cumulant tensors.
     We show how to detect the transition from
    Gaussian distributed data to non-Gaussian ones in a~data stream. In order to
    achieve high implementation efficiency of operations on super-symmetric
    tensors, such as cumulant tensors, we employ the block structure to store
    and calculate only one hyper-pyramid part of such tensors. 
\end{abstract}

\maketitle

\section{Introduction}
Cumulants of order one and two of $n$-dimensional multivariate data, \ie{} mean
vector and covariance matrix are widely used in signal and data processing, for
example, in one of the most widely used algorithm in data and signal processing,
namely Principal Component Analysis. Cumulants of order one and two describe
completely statistically signal or data whose values are govern by a~Gaussian
distribution. In many real-life cases data or signals are not normally
distributed. In this case it is necessary to employ higher order cumulants, such
as, for example, skewness and kurtosis, to analyze this kind of data.

As the high order cumulant of $n$ dimensional multivariate data we understand
the super-symmetric\footnote{A tensor is super-symmetric if it is invariant
under permutation of its indices.}, cumulant tensor $\tensor{C} \in \Real^{[n,
d]}$ of $d \geq 3$ modes, each of size $n$. Importantly they are zeros only if
calculated for data sampled from multivariate Gaussian distribution
\cite{kendall1946advanced,lukacs1970characteristics}. High order cumulants carry
information about the divergence of the empirical distribution from the
multivariate Gaussian one, hence we use them to extract such information from 
data.

Calculation of higher order cumulants for multi-dimensional data is time
consuming. Furthermore, such data are often recorded in form of a~stream and
hence the on-line scheme of calculation and updates of cumulants is useful to
analyze them. In this paper we present an~efficient algorithm for calculation of
cumulants of arbitrary order in a sliding window for data streams. We show the
application of this algorithm to detect change in the underling distribution of
multivariate time-series. Our algorithm uses so called block-structure, which is
a~data structure designed for efficient storage and processing of symmetric
tensors.

\subsection{Motivation}
Our motivation to design such an algorithm comes from the fact that there
exist many contemporary applications of higher order cumulants based algorithms
in data processing. Typically these algorithms employ cumulants up to order
four, and rarely up to order six. This limitation comes mainly from two factors:
high computational cost of calculating higher order cumulants, and large amounts
of data samples required to estimate faithfully higher order cumulants. Nowadays
computational power is widely available and amounts of data collected every day
is increasing dramatically. Therefore we believe that algorithms requiring usage
of high order cumulants will be employed more widely in the near future. Yet, as
it was pointed out by \cite{stefanowsk2017exploring}, processing data streams is
a challenging task because it imposes constraints on memory usage, processing
time, and number of data inputs reads. The algorithm presented in this work is
dedicated to process efficiently on-line large data streams.

High order cumulants are used to analyze signal data, such as audio signals, for
example in direction-finding methods of the multi-source signal (q-MUSIC
algorithm) \cite{chevalier2006high}. Additionally, high order cumulants are
being used in signal filtering problems
\cite{geng2011research,latimer2003cumulant} or neuroimaging signals
analysis~\cite{birot2011localization,becker2014eeg}. The neuroscience
application often uses the Independent Components Analysis (ICA)
\cite{hyvarinen2013independent}, that can be evaluated by means of high order
cumulant tensors~\cite{blaschke2004cubica,virta2015joint}. Another important
issue that requires fast algorithm to compute and update high order cumulants is
financial data analysis, especially concerning high frequency financial data,
where we deal with large data sets and the computational time is a crucial
factor. For multi-assets portfolio analysis, high order cumulant tensors
measures risk \cite{rubinstein2006multi,martin2013consumption}, especially
during a crisis where large fluctuations of assets values are possible
\cite{arismendi2014monte,jondeau2015moment,domino2016usecum}.

While estimating high order statistics from data, there rises a problem of high
estimation error. In general, large data set required for the accurate
estimation of high order cumulants from data. This is discussed in
\cite{domino2017tensorsnet} in some details. Unfortunately, large data set
requires large computational time what becomes problematic if we want to analyze
$n$-variate data on-line and $n$ is respectively large. To solve this problem we
introduce an algorithm, that computes high order statistics in a sliding window of length $t$.
Statistics are updated every time a new data batch of size $t_{\mathrm{up}}$ is collected.

The values of parameters $t$ and $t_{\mathrm{up}}$ depend on particular
application. On one hand $t$ and $t_{\mathrm{up}}$ have to be large enough for
an accurate approximation of the statistics, on the other hand the larger they
are the weaker the time resolution of accessible for an application. We
typically choose $t_{\mathrm{up}} = \alpha t$ with $\alpha = 2.5\%, 5\%, \ldots$
Given such parameters, we have reached over an order of magnitude speedup
compared with a simple cumulants' recalculation using fast algorithm introduced
in \cite{domino2017tensorsnet}. In both cases we use the block structure
\cite{schatz2014exploiting} that allows to calculate and store efficiently
super-symmetric cumulant and moment tensors \cite{domino2017tensorsnet}. We show
that using the presented algorithm we can analyze data recorded at frequencies
up to $2000$ Hz from $150$ Hz, depending in the number of marginal variables
$n$: $60$---for the higher frequency figure to $120$---for the lower figure, on
a~modern six-core workstation.

\subsection{Paper structure}
The paper is organized as follows. In Section \ref{s::statistics} we present
formulas and algorithm employed to calculate cumulants of a data stream, the
input data format, the sliding window mechanism, the block structure, moments
tensors updates, cumulants calculation, and the complexity analysis. In
Section~\ref{s::implementation} we discuss the algorithm implementation in Julia
programming language, and the performance tests of the implementation. In
Section~\ref{s::exemplaryapp} we introduce an illustratory application of our
algorithm applied to analyze on-line the statistics of a~data stream and
analyze the maximal frequency of data that can be calculated on-line given
a~computer hardware.

\section{Statistics updates}\label{s::statistics}
\subsection{Data format}\label{s::dataformat}
Let us consider the data that consists of $t$ realizations sampled from $n$-dimensional
multivariate distribution forming an observation window whose number we will index by $w$:
\begin{equation}\label{eq::variable}
\mathbb{R}^{t \times n} \ni \mat{X}^{(w)} = 
\begin{bmatrix}
x_{1,1}^{(w)} & \dots & x_{1,n}^{(w)}  \\ 
\vdots & \ddots & \vdots \\ 
x_{t,1}^{(w)} & \dots & x_{t,n}^{(w)}  \\ 
\end{bmatrix}.
\end{equation}
Note that samples form rows in the data matrix.

\begin{figure}
\centering
\begin{tikzpicture}
\matrix (m) [matrix of nodes, minimum width=1.cm, align=center, minimum height=1.cm, 
			 column sep=0.3cm, 
			 every node/.style={anchor=base,text depth=.5ex,text height=2ex,text width=1cm}]  
{
	{} 						&	{$\mat{X}^{(1)}_{(-)}$} &	{}  \\
	{$\mat{X}^{(1)}$} 		&	{} 						&	{$\mat{X}^{(2)}_{(-)}$}  \\
	{} 						&	{$\mat{X}^{(2)}$} 		&	{}  \\
	{$\mat{X}^{(1)}_{(+)}$}	&	{} 						&	{$\mat{X}^{(3)}$}  \\
	{} 						&	{$\mat{X}^{(2)}_{(+)}$}	&	{}  \\
	{} 						&	{}						&	{$\mat{X}^{(3)}_{(+)}$}  \\
};
\draw (m-1-1.north west) rectangle (m-3-1.south east);
\draw [dashed] (m-4-1.north west) -- (m-4-1.south west) -- (m-4-1.south east) -- (m-4-1.north east);
\draw [dashed] (m-1-2.south west) -- (m-1-2.north west) -- (m-1-2.north east) -- (m-1-2.south east);
\draw (m-2-2.north west) rectangle (m-4-2.south east);
\draw [dashed] (m-5-2.north west) -- (m-5-2.south west) -- (m-5-2.south east) -- (m-5-2.north east);
\draw [dashed] (m-2-3.south west) -- (m-2-3.north west) -- (m-2-3.north east) -- (m-2-3.south east);
\draw (m-3-3.north west) rectangle (m-5-3.south east);
\draw [dashed] (m-6-3.north west) -- (m-6-3.south west) -- (m-6-3.south east) -- (m-6-3.north east);
\draw [->] ($(m-1-1.north west) - (0.2cm,0)$) -- node [left] {time}  ($(m-6-1.south west) - (0.2cm,0)$) ;
\end{tikzpicture}
\caption{Schematic representation of data flow in sliding window mechanism. In
the picture the time flows from top to bottom. Subsequent windows are placed
from left to right. Each multivariate data sample forms a row of a matrix.}
\label{fig::slidingwindow}
\end{figure}
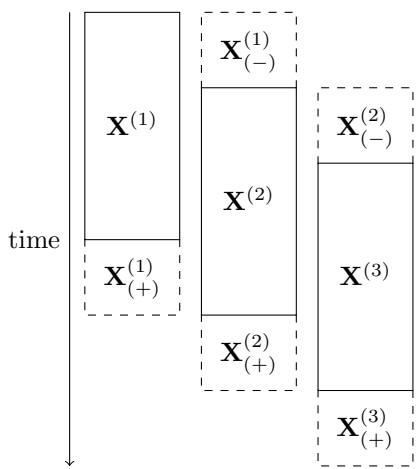

Further consider an update, that consists of another $t_{\mathrm{up}}$ $n$-dimensional realizations:
\begin{equation}\label{eq::upvariable}
\mathbb{R}^{t_{\mathrm{up}} \times n} \ni \mat{X}_{(+)}^{(w)} = 
\begin{bmatrix}
x_{t+1,1}^{(w)} & \dots & x_{t+1,n}^{(w)}  \\ 
\vdots & \ddots & \vdots \\ 
x_{t+t_{\mathrm{up}},1}^{(w)} & \dots & x_{t+t_{\mathrm{up}},n}^{(w)}  \\ 
\end{bmatrix},
\end{equation}
that will be concatenated to $\mat{X}^{(w)}$ in order to form a~new window.
Additionally forming of a~new window will require to drop first
$t_{\mathrm{up}}$ realizations represented by the following matrix
\begin{equation}\label{eq::bisvariable}
\mathbb{R}^{t_{\mathrm{up}} \times n} \ni \mat{X}_{(-)}^{(w)} = 
\begin{bmatrix}
x_{1,1}^{(w)} & \dots & x_{1,n}^{(w)}  \\ 
\vdots & \ddots & \vdots \\ 
x_{t_{\mathrm{up}},1}^{(w)} & \dots & x_{t_{\mathrm{up}},n}^{(w)}  \\ 
\end{bmatrix}.
\end{equation}
The new observation window $w+1$ is given by the following equation
\begin{equation}\label{eq::uvariable}
\begin{split}
\mathbb{R}^{t \times n} \ni \mat{X}^{(w+1)} 
& =
\begin{bmatrix}
x_{t_{\mathrm{up}}+1,1}^{(w)} & \dots & x_{t_{\mathrm{up}}+1,n}^{(w)}  \\
\vdots & \ddots & \vdots \\ 
x_{t+t_{\mathrm{up}},1}^{(w)} & \dots & x_{t+t_{\mathrm{up}},n}^{(w)}  \\
\end{bmatrix} =\\
& =
\begin{bmatrix}
x_{1,1}^{(w+1)} & \dots & x_{1,n}^{(w+1)}  \\
\vdots & \ddots & \vdots \\ 
x_{t,1}^{(w+1)} & \dots & x_{t,n}^{(w+1)}  \\
\end{bmatrix}.
\end{split}
\end{equation}
The sliding window mechanism is visualized in Fig.~\ref{fig::slidingwindow}.

\subsection{Sliding window}\label{s::slidingwindow}
The algorithm presented in this work calculates cumulants of a data stream 
in a~sliding window. It is assumed that data arrive continuously to a~system and
are fed to the algorithm in typically small batches. The algorithm uses only
a~subset of current data stored in a buffer and minimal required statistics. As
new data are incoming the calculations are performed on stored data and
statistics. Historical data are iteratively discarded. The main loop is
summarized in Algorithm~\ref{alg::mainloop}. Which consists of the following
steps: acquire new batch of data; calculate the oldest batch of data; update
moments; calculate cumulants; update the data buffer.
\begin{algorithm*}[h!]
	\caption{Sliding window cumulant calculation algorithm}\label{alg::mainloop}
	\begin{algorithmic}[1]
		\State \textbf{Input}: $\mat{X}^{(1)}$ -- first data batch
		\State \textbf{Output:} $\tensor{C}_1(\mat{X}^{(w)}), \ldots, 
		\tensor{C}_d(\mat{X}^{(w)})$ -- cumulants for windows $1:w$
		\State Calculate moments $\tensor{M}_1(\mat{X}^{(1)}), \ldots, 
		\tensor{M}_d(\mat{X}^{(1)})$
		\State $w\gets 1$
		\For{$w$}
		\State Acquire $\mat{X}^{(w)}_{(+)}$.
		\State Calculate $\mat{X}^{(w)}_{(-)}$ from fist $t_{\mathrm{up}}$ 
		rows of $\mat{X}^{(w)}$.
		\State $\tensor{M}_1(\mat{X}^{(w+1)}), \ldots, 
		\tensor{M}_d(\mat{X}^{(w+1)}) 
		\gets \text{momentsupdate}(\tensor{M}_1(\mat{X}^{(w)}), 
		\ldots,\tensor{M}_d(\mat{X}^{(w)}), \mat{X}^{(w)}_{(+)}, 
		\mat{X}_{(-)}^{(w)})$
		\State $\tensor{C}_1(\mat{X}^{(w+1)}), \ldots, 
		\tensor{C}_d(\mat{X}^{(w+1)}) \gets 
		\text{mom2cums}(\tensor{M}_1(\mat{X}^{(w+1)}), 
		\ldots, 
		\tensor{M}_d(\mat{X}^{(w+1)}))$
		\State Calculate $\mat{X}^{(w+1)}$ by concatenating row by row 
		$\mat{X}^{(w)}$ with $\mat{X}^{(w)}_{(+)}$ and remove rows 
		belonging to $\mat{X}^{(w)}_{(-)}$.
		\State Emit $\tensor{C}_1(\mat{X}^{(w+1)}), \ldots, 
		\tensor{C}_d(\mat{X}^{(w+1)})$. \label{ln::emitcumulants}
		\State $w\gets w+1$
		\EndFor
	\end{algorithmic}
\end{algorithm*}  

\subsection{Block structure}\label{ss::blockstructure}
Moments and cumulants are super-symmetric tensors, therefore we use block
structure as introduced in \cite{schatz2014exploiting} to compute and store them
effectively. Using the such a~block structure we store and compute only one
hyper-pyramidal part of the super-symmetric tensor in blocks of size $b^d$,
where $b$ is a~parameter of the storage method. One advantage of the block
structure is that it allows for efficient further processing of cumulants what 
was discussed in~\cite{domino2017tensorsnet}.

\subsection{Moment tensor updates}\label{s::momup}
Given data $\mat{X}\in \Real^{t\times n}$ the super-symmetric moment tensor of 
order $d$:
$\mathcal{M}_d(\mat{X}) \in \Real^{[n,d]}$ consists of the following elements:
\begin{equation}{\label{eq:mom}}
    m_{\midx{i}}(\mat{X}) = \frac{1}{t} \sum_{l = 1}^t 
    \left(\prod_{i_k \in \midx{i}}
    x_{l,i_k} 
    \right).
\end{equation}
where $\midx{i} = (i_1, \ldots, i_d)$ is element's multi-index and $i_1, \ldots,
i_d \in 1:n$. A naive approach to calculate moments
$\tensor{M}\left(\mat{X}^{(w)}\right)$ would be to calculate all
$m_{\midx{i}}(\mat{X}^{(w)})$ for each window $w$.
But in order to reduce the amount of computation required to calculate moments in sliding windows
we take advantage of the fact, that given $\mat{X}_{(-)}^{(w)}$ and $\mat{X}_{(+)}^{(w)}$ it
is easy to update each element of the moment tensor using the following relation:
\begin{equation}\label{eq::momupdate}
\begin{split}
&m_{\midx{i}}\left(\mat{X}^{(w+1)}\right) = \frac{1}{t} \sum_{l = 1+t_{\mathrm{up}}}^{t+t_{\mathrm{up}}} 
\left(
\prod_{i_k \in \midx{i}} x_{l,i_k}^{(w)} \right) = \\ 
&=\frac{1}{t}\sum_{l = 1}^{t}\left(\prod_{i_k \in \midx{i}} x_{l,i_k}^{(w)} \right)+\\
&+ \frac{t_{\mathrm{up}}}{t} 
\left(\frac{1}{t_{\mathrm{up}}}\sum_{l = 1+t}^{t+t_{\mathrm{up}}}\left(\prod_{i_k \in \midx{i}} x_{l,i_k}^{(w)}  \right) - \frac{1}{t_{\mathrm{up}}}\sum_{l = 1}^{t_{\mathrm{up}}}\left(\prod_{i_k \in \midx{i}} x_{l,i_k}^{(w)} \right) 
\right) =\\
&= m_{\midx{i}}\left(\mat{X}^{(w)}\right) + \frac{t_{\mathrm{up}}}{t} \left(m_{\midx{i}}\left(\mat{X}_{(+)}^{(w)}\right)- m_{\midx{i}}\left(\mat{X}_{(-)}^{(w)}\right)\right).
\end{split}
\end{equation}
We can write Eq.~\eqref{eq::momupdate} using 
tensor notation in the following tensor form:
\begin{equation}\label{eq::momupdatetensor}
\begin{split}
\tensor{M}\left(\mat{X}^{(w+1)}\right) = & \tensor{M}\left(\mat{X}^{(w)}\right) +  \\
+&\frac{t_{\mathrm{up}}}{t}\left(\tensor{M}\left(\mat{X}^{(w)}_{(+)}\right) - \tensor{M}\left(\mat{X}^{(w)}_{(-)}\right) \right).
\end{split}
\end{equation}
\begin{algorithm*}
	\caption{\fct{momentsupdate}}\label{alg::momentsupdate}
	\begin{algorithmic}[1]    
		\State \textbf{Input}: data -- $\mat{X}_{(+)}^{(w)} \in 
		\Real^{t_{up}\times n}$, 
		$\mat{X}_{(-)}^{(w)} \in \Real^{t_{up}\times n}$, moments -- 
		$\tensor{M}_1(\mat{X}^{(w)}), 
		\ldots, 
		\tensor{M}_d(\mat{X}^{(w)})$.
		\State \textbf{Output:} updated moments -- 
		$\tensor{M}_1(\mat{X}^{(w+1)}), \ldots, 
		\tensor{M}_d(\mat{X}^{(w+1)})$
		\For{$s \gets 1:d$}
		\State
		$
		\tensor{M}_s(\mat{X}^{(w+1)}) \gets \tensor{M}_s(\mat{X}^{(w)}) + 
		\frac{t_{\mathrm{up}}}{t}\left(\tensor{M}\left(\mat{X}^{(w)}_{(+)}\right)
		 - \tensor{M}\left(\mat{X}^{(w)}_{(-)}\right) \right)
		$
		\EndFor 
		\Comment{See Eq.~\eqref{eq::momupdatetensor}} 
		\State \Return $\tensor{M}_1(\mat{X}^{(w+1)}), \ldots, 
		\tensor{M}_d(\mat{X}^{(w+1)})$
	\end{algorithmic}
\end{algorithm*}  
Exploiting this form we can write the Algorithm \ref{alg::momentsupdate} which
calculates moments in a sliding window $w+1$ given moments of window $w$, and the data batches
$\mat{X}_{(-)}^{(w)}$ and $\mat{X}_{(+)}^{(w)}$.

There exists a~different approach to this problem. We could calculate
${t}/{t_{\mathrm{up}}}$ moments 
of data batches and organize the moments in a~FIFO queue 
\begin{equation}\left(\tensor{M}\left(\mat{X}_{(+)}^{(w_1)}\right), 
\tensor{M}\left(\mat{X}_{(+)}^{(w_2)}\right), 
\ldots,
\tensor{M}\left(\mat{X}_{(+)}^{(w_{{t}/{t_{\mathrm{up}}}})}\right)
\right).
\end{equation} 
With arrival of new batch its moments would be calculated and added to the
aggregate moments then the oldest batch moments would be subtracted from the
aggregate. This scheme reduces the amount of calculations because it does not
require to calculate moments of $\mat{X}_{(-)}^{(w)}$ for each window $w$, but
requires the storage of ${t}/{t_{\mathrm{up}}}$ moments for data
$\mat{X}_{(+)}^{(w_b)}$ for $w_b \in w : w + {t}/{t_{\mathrm{up}}}$. Therefore
in approach we propose we have traded some of the computational complexity for
the reduction in memory size requirements.

The moment tensor computation and storage in the block structure is explained
in details in \cite{domino2017tensorsnet} from where we can conclude that if $b
\ll n$ and $d \ll n$ we need approximately $\frac{n^d}{d!} (d-1) t$
multiplications to compute $\tensor{M}_d(\mat{X})$. Analogically, we need
$\frac{n^d}{d!} (d-1) t_{\mathrm{up}}$ multiplications to compute
$\tensor{M}\left(\mat{X}_{(+)}^{(w)}\right)$ and the same number of
multiplications to compute $\tensor{M}\left(\mat{X}_{(-)}^{(w)}\right)$. Obviously a~simple
recalculation of $\tensor{M}(\mat{X}^{(w+1)})$ would require $\frac{n^d}{d!} 
(d-1) t$
multiplications. Hence given $\tensor{M}\left(\mat{X}^{(w)}\right)$ computed priorly, the
theoretical speedup factor of the update compared with a simple recalculation
of $\tensor{M}\left(\mat{X}^{(w+1)}\right)$ would be:
\begin{equation}\label{eq::upcmplx}
\frac{\frac{n^d}{d!} (d-1) t}{\frac{n^d}{d!} (d-1) t_{\mathrm{up}} + \frac{n^d}{d!} 
(d-1) t_{\mathrm{up}}} = \frac{t}{2 t_{\mathrm{up}}},
\end{equation}
what is significant especially if $t_{\mathrm{up}} \ll t$. In next two subsections we 
are going to show how to use the moment update scheme to update cumulant 
tensors.

\subsection{Cumulant updates calculation}\label{s::cumcalc}
Given the moment tensor update scheme, due to the recursive relation between
moments and cumulants tensors \cite{barndorff1989asymptotic}, we can use this
scheme to form a cumulants' update algorithm. The recursive relation between
cumulants and moments was discussed in details the previous
work~\cite{domino2017tensorsnet}. Here this relation is summarized in a form of
Algorithm~\ref{alg::cumulants}. This algorithm calculates cumulants' tensors
$\tensor{C}_s(\mathbf{X})\in \mathbb{R}^{[n,s]}$ for orders $s\in\{1,2,\ldots,
d\}$, given moments $\tensor{M}_1(\mat{X}), \ldots,\tensor{M}_d(\mat{X})$.

\begin{algorithm}[h]
	\caption{\fct{moms2cums}}
	\label{alg::cumulants}
	\begin{algorithmic}[1]  
		\State \textbf{Input}: $\tensor{M}_1(\mat{X}), 
		\ldots,\tensor{M}_d(\mat{X})$ -- moments
		\State \textbf{Output:} $\tensor{C}_1(\mat{X}), \ldots, 
		\tensor{C}_d(\mat{X})$ -- cumulants
		\For{$s \gets 1:d$}
		\State $\tensor{C}_s(\mat{X}) \gets \tensor{M}_s(\mat{X}) - \tensor{A}$ 
		\Comment{Calculate elements of $\tensor{A}$ using algorithm 
			\ref{alg::symoutprod}}
		\EndFor 
		\State \Return $\tensor{C}_1(\mat{X}), \ldots, \tensor{C}_d(\mat{X})$
	\end{algorithmic}
\end{algorithm}    

\begin{algorithm}[h]
	\caption{Calculation of symmetrized outer product}\label{alg::symoutprod}
	\begin{algorithmic}[1]  
		\State \textbf{Input}: $\midx{i}$ -- multi-index of cumulant tensor, 
		$s$ -- order of cumulant being calculated,
		$\tensor{C}_{1}(\mat{X})$, $\tensor{C}_{2}(\mat{X})$, \ldots,  
		$\tensor{C}_{s-1}(\mat{X})$ -- cumulant tensors of lower orders
		\State \textbf{Output:} $\tensor{A}_{\midx{i}}$ -- element of 
		super-symmetric tensor $\tensor{A}$
		\State $\tensor{A}_{\midx{i}}=0$
		\For{$\sigma \gets 2:s$} \label{ln::sigmaloop}
		\State calculate partitions of the set $1:s$ into $\sigma$ parts 
		\Comment{using Knuth's algorithm \cite[Section 7.2.1.4]{knuth2011art}}
		\For{$\xi \in \text{partitions}$} \label{ln::loopparts}
		\State $\mathrm{a} \gets 1$
		\For{$k \in \xi$}
		\State $\mathrm{a}\gets\mathrm{a}\times 
		\tensor{C}_{\midx{i}(k)}(\mat{X})$ \label{ln::multiply}
		\EndFor
		\State $\tensor{A}_{\midx{i}} \gets \tensor{A}_{\midx{i}} + 
		\mathrm{a}$
		\EndFor
		\EndFor
	\end{algorithmic}
\end{algorithm}   

\subsection{Complexity analysis}\label{s::complexityanalysis}
Despite using
cumulants--moment recursive relation form \cite{domino2017tensorsnet}, there
are important computational differences between the cumulants' updated scheme
proposed in this paper and cumulants calculation scheme proposed in
\cite{domino2017tensorsnet}, see Eq.~(34) therein. In the first case:
\begin{enumerate}
    \item we need much less arithmetic operations to update a moment tensor than in the second case since 
    $t_{\mathrm{up}} \ll t$, but
    \item we need slightly more computational power to compute $\tensor{A}$ in
    Algorithm~\ref{alg::symoutprod}. In the first case we can not use central moments 
    for $\tensor{M}_d$ because in general updates affect the centering of the data.
    Hence in the first case the inner loop starting in line~\ref{ln::loopparts} of 
    Algorithm~\ref{alg::symoutprod} runs over all partitions, in contrary to the second case, 
    where similar algorithm sums over partitions containing only elements of size $\geq 2$.
\end{enumerate}

%Let us now move to an example and number of multiplications involved in 
%Eq~\eqref{eq::cum_rec}.
%\begin{example}
%    \begin{equation}
%    \begin{split}
%        \{[P_{2}(1 : 3)]\} &= \{\{(1,2), (3)\}, \ \{(1,3), (2)\}, \ \{(2,3), 
%        (1)\}\}, \\
%        \{[P_{3}(1 : 3)]\} &= \{\{(1), (2) ,(3)\}.
%    \end{split}
%    \end{equation}
%    hence $3$\textsuperscript{th} cumulant in elementwise notation
%    \begin{equation}
%    \begin{split}
%    c_{(i_1, i_2, i_3)}(\textbf{X}) &= m_{(i_1, 
%        i_2, i_3)}(\textbf{X}) - c_{(i_1, i_2)}(\textbf{X})\cdot 
%    c_{(i_3)}(\textbf{X}) - c_{(i_1, i_3)}(\textbf{X})\cdot 
%    c_{(i_2)}(\textbf{X}) \\ &- c_{(i_2, i_3)}(\textbf{X})\cdot 
%    c_{(i_1)}(\textbf{X}) - c_{(i_1)}(\textbf{X}) \cdot 
%    c_{(i_2)}(\textbf{X}) \cdot c_{(i_3)}(\textbf{X}).
%    \end{split}
%    \end{equation}    
%\end{example}
In order to analyze the computational complexity of sliding window cumulant
calculation algorithm, we have to count the number of multiplications performed
in line~\ref{ln::multiply} of the Algorithm~\ref{alg::symoutprod}. This number
is given by:
\begin{equation}{\label{eq::n1}}
\begin{split}
\sum_{\sigma = 1}^{d}S(d, \sigma)(\sigma - 1) = & \\ = \sum_{\sigma = 2}^{d}S(d, \sigma)(\sigma - 1)  \leq & (d-1)\sum_{\sigma = 2}^{d}S(d, \sigma) < \\ < & (d-1)B(d),
\end{split}
\end{equation}
where $S(d, \sigma) > 0$
is the number of partitions of set of size $d$
into $\sigma$ parts, \ie~the Stirling Number of the second kind \cite{grahamconcrete};
the sum $\sum_{\sigma = 1}^{d} S(d, \sigma) = B(d)$,
is the Bell number \cite{comtet1974advanced}, the number of all
partitions of the set of size $d$. 
The upper limit $(d-1)B(d)$ will be used further to approximate the
number of multiplications required. 

The number of multiplications is reduced due to the use of the block
storage of super-symmetric tensors. We need only to calculate approximately
$\frac{n^d}{d!}$ tensor elements see~\cite{domino2017tensorsnet}.
Given a moment tensor $\tensor{M}_d$
and cumulant tensors $\tensor{C}_1, \ldots, \tensor{C}_{d-1}$ we can approximate
number of multiplications to compute $\tensor{C}_d$ by
\begin{equation}\label{eq::cumcomplex}
\frac{n^d}{d!}(d-1)B(d).
\end{equation}
Nevertheless it is important to notice, that there is some additional computational 
overhead in the implementation due to operations on relatively small blocks.

Referring to Eq.~\eqref{eq::upcmplx} in order to update a series of moments we need 
approximately:
\begin{equation}
\#N_{\mathrm{mup}}(d) \approx \sum_{k = 1}^d 2\frac{n^k}{k!} (k-1) t_{\mathrm{up}}
\end{equation}
multiplications. Further according to Eq.~\eqref{eq::cumcomplex} to compute a 
series of cumulant tensors given a series of moment tensors we need 
approximately:
\begin{equation}
\sum_{k = 1}^d \frac{n^k}{k!}(k-1)B(k)
\end{equation}
multiplications. Finally to update a series of cumulants we need 
\begin{equation}{\label{eq:complx}}
\begin{split}
\#N_{\mathrm{cup}} & \approx \sum_{k = 1}^d 2\frac{n^k}{k!} (k-1) t_{\mathrm{up}} +  \sum_{k = 1}^d  \frac{n^k}{k!}(k-1)B(k)=\\ & = 
\sum_{k = 1}^d 2\frac{n^k}{k!} (k-1) (2 t_{\mathrm{up}} + B(k))
\end{split}
\end{equation}
multiplications. In practice, we use cumulants' orders of $d = 4,5,6$, number of
data $t > 10^5$, batch size of $t_{\mathrm{up}} = \alpha t$, where $\alpha =
2.5\%, 5\%, \ldots$, and number of variables $n \gg d$. Further given that
Bell number $B(d)$ rises rapidly with $d$ \cite{comtet1974advanced} the last 
term of the sum in Eq.~\eqref{eq:complx} is dominant 
and hence the final number of multiplications can be approximated by:
\begin{equation}{\label{eq::complxn}}
\#N_{cup} \approx \frac{n^d}{d!} (d-1) (2 t_{\mathrm{up}} + 
B(d)).
\end{equation}
Simple cumulant series recalculation using \cite{domino2017tensorsnet} requires 
approximately $\sum_{k = 1}^d \frac{n^k}{k!} (k-1) t \approx \frac{n^d}{d!} 
(d-1) t$ multiplications. The final speedup 
factor compared with such recalculation is:
\begin{equation}
\frac{t}{2t_{\mathrm{up}}+B(d)}.
\end{equation}
The first term in the denominator corresponds to 
Algorithm~\ref{alg::momentsupdate}, while the 
second one to Algorithm~\ref{alg::cumulants}. For analyzed parameters' values 
the function \fct{momentsupdate} is 
more computationally costly in comparison with \fct{moms2cums} by a factor of 
$\frac{2t_{\mathrm{up}}}{B(d)}$. For example for $t_{\mathrm{up}} = 25000$ and 
$d = 4$ the this factors is of three orders of magnitude.

In the following Sections we present the computer implementation of the 
cumulants 
updates algorithm in \proglang{Julia} programming language, and performance tests.

\section{Implementation and performance}\label{s::implementation}
\subsection{Implementation}
The sliding window cumulant calculation algorithm was implemented in \proglang{Julia} programming 
language 
\cite{bezanson2012julia,bezanson2014julia,bezanson2014array} and provided on 
the Zenodo repository \cite{cup}. Our implementation uses the block structure 
provided in \cite{st}, and parallel computation via the 
function \fct{pmap} implemented in \proglang{Julia} programming language. 
For the parallel computation implementation we perform the following.
\begin{enumerate}
    \item We use parallel implementation of moment tensor calculation 
    introduced in \cite{domino2017tensorsnet}, \ie \ data are split into $p$ 
    non-overlapping sub-series, where $p$ is a 
    number of workers, next we compute moment tensors for each sub-series and 
    combine them into a single moment tensor. 
    \item We have also parallelized the for loop in line~\ref{ln::sigmaloop}
    of Algorithm~\ref{alg::symoutprod} using 
    \fct{pmap} function which is one of the ways \proglang{Julia}
    programming language implements a parallel for. The advantage of this
    solution is that each term of that sum is super-symmetric and we can compute
    it using block structure. The disadvantage is that the sum has only $d-1$
    elements hence for large number of workers we do not take full advantage of
    the parallel implementation.
\end{enumerate}
Despite some inefficiencies of parallel implementation, we obtain large speedup 
due to multiprocessing what is presented below.

\subsection{Performance tests}{\label{s::tests}}
In what follows we present performance tests carried out mainly using
multiple CPU cores. All tests were
performed on a~computer equipped with \texttt{Intel(R) Core(TM) i7-6800K CPU @
3.40GHz} processor providing $6$ physical cores and $12$ computing cores
with hyper-threading, and $64$ GB of random access memory.

We start with determining the optimal block size parameter $b$ of the block
structure (see Subsection~\ref{ss::blockstructure}). This parameter has complex impact on the computational time. On the
one hand the higher $b$, the more computational and storage redundancy while
calculating moment tensors, due to larger diagonal blocks.
On the other hand the lower $b$ the more computational overhead due to larger amount of
operations performed on small blocks.

\begin{figure}[t!]
    \centering
    \subfigure[$d = 4, n = 60$\label{fig::b4}]
    {\includegraphics[width=0.4\textwidth]{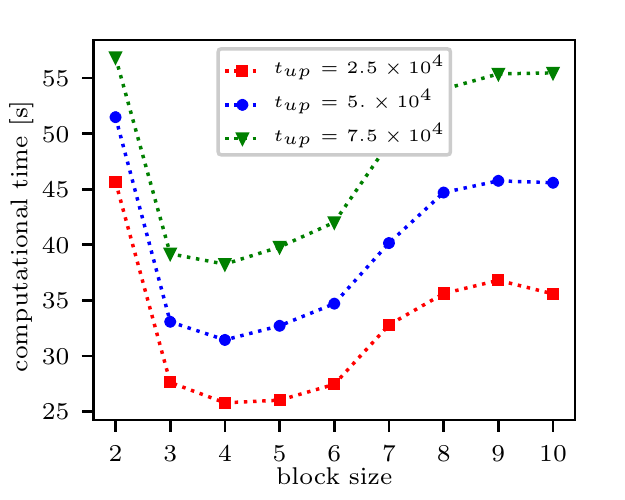}}
    \subfigure[$d = 4, n = 120$\label{fig::b4120}]
    {\includegraphics[width=0.4\textwidth]{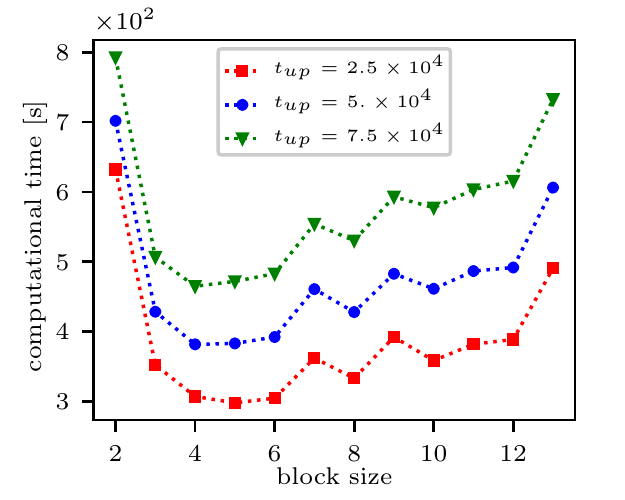}}
    \subfigure[$d = 5, n = 60$\label{fig::b5}]
    {\includegraphics[width=0.4\textwidth]{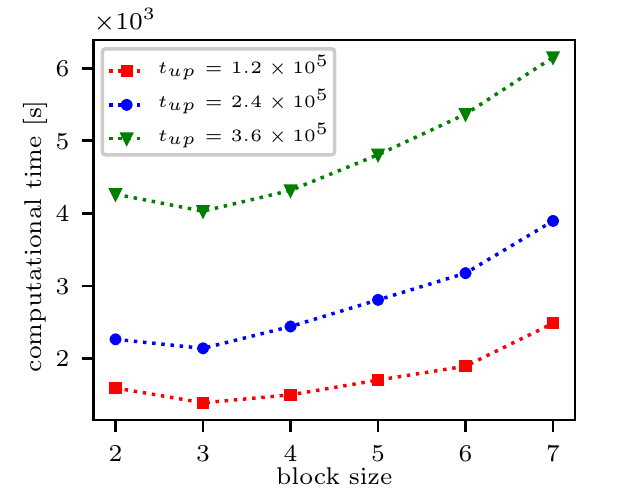}}
    \caption{Cumulants updates computational times for different block sizes 
    $b$ and the multiprocessing implementation on $6$ workers.}
    \label{fig::cb}
\end{figure}
In Figure~\ref{fig::cb} we present the computational time of the update of
cumulant tensors series of order $1, \ldots, d$ for different block sizes. One
can observe that, the higher cumulant order $d$ the lower optimal block size $b$.

Finally fluctuations of computational time vs. block size are caused by the fact
that, in our implementation, if $b$ does not divide $n$ some blocks are not
hyper-squares and hence calculation of their size and block sizes conversion
cause additional computational overhead. The computations of the optimal block
size were performed using $6$ parallel worker processes.

\begin{figure}[t]
    \centering
    \subfigure[$d = 4, n = 120, b = 4$\label{fig::p4120}]%
        {\includegraphics[width=0.4\textwidth]{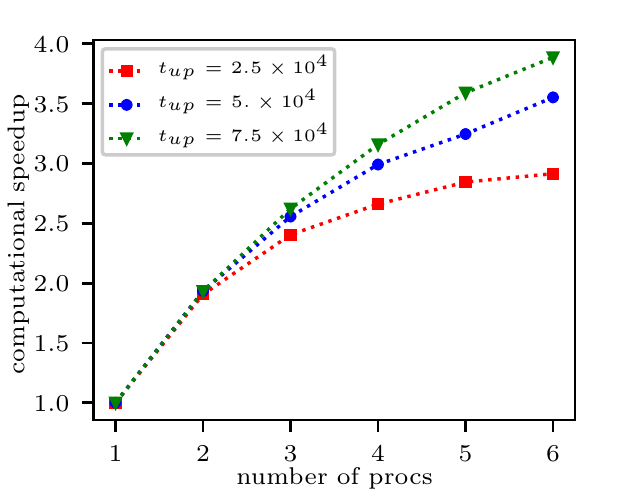}}
    \subfigure[$d = 5, n = 60, b = 3 $\label{fig::p5}]%
        {\includegraphics[width=0.4\textwidth]{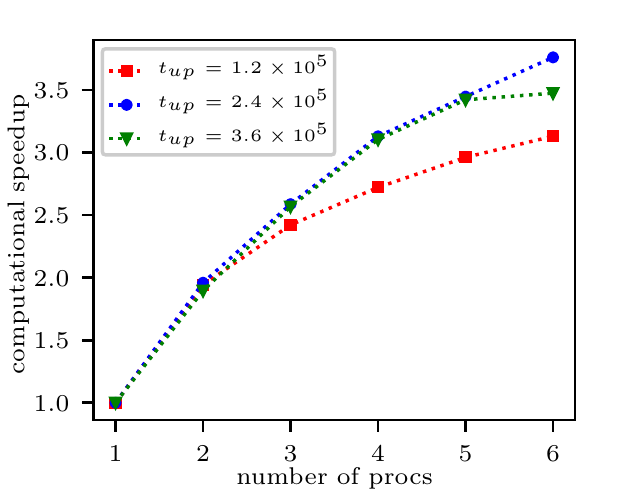}}
    \caption{\label{fig::p} Performance tests for multiprocessing implementation.}
\end{figure}

Scalability of the algorithm with raising number of CPU cores is pretended in
Figure~\ref{fig::p}.
At first the computational time speedup is proportional to number of workers 
as it should be expected, however for large number of workers 
we do not fully take advantage of parallel implementation what is discussed in 
a previous section. Despite this problem we have still large speedup due to use 
multiple cores.

\begin{figure*}[t]
    \centering
    \subfigure[$t = 5\times 10^5, d = 4, b = 4$.\label{fig::c4_05}]%
        {\includegraphics[width=0.4\textwidth]{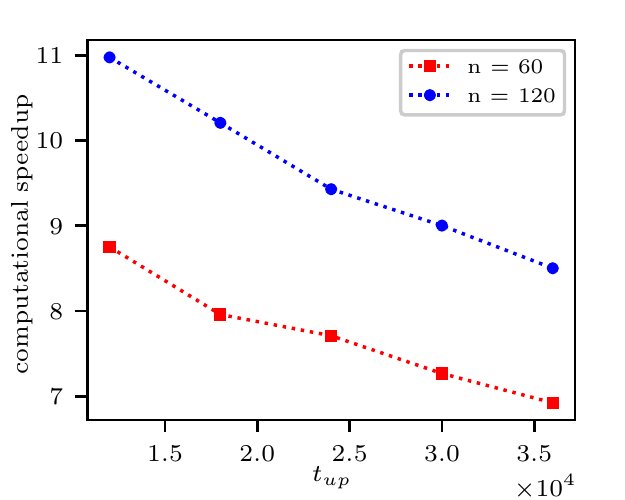}}
    \subfigure[$t = 1\times 10^6, d = 4, b = 4$.\label{fig::c4_2}]%
        {\includegraphics[width=0.4\textwidth]{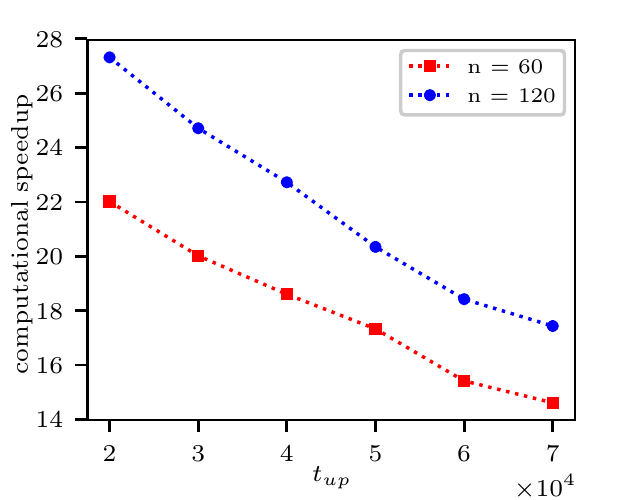}} 
    \\ 
    \subfigure[$t = 5 \times 10^6, d = 5, b = 3$.\label{fig::c5}]%
        {\includegraphics[width=0.4\textwidth]{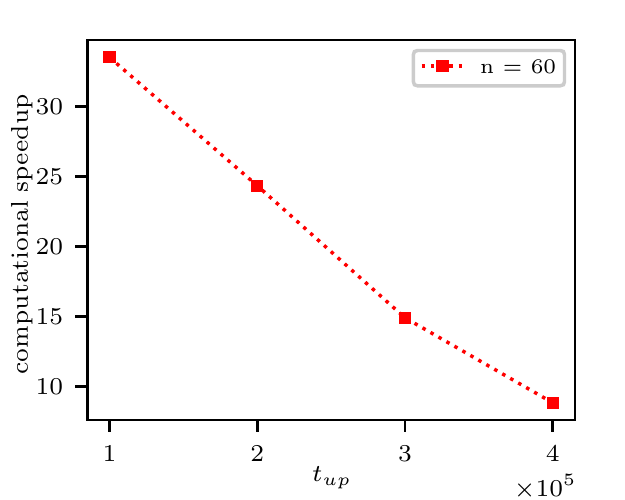}}
    \subfigure[$t = 25 \times 10^6, d = 6, b = 2$.\label{fig::c6}]%
        {\includegraphics[width=0.4\textwidth]{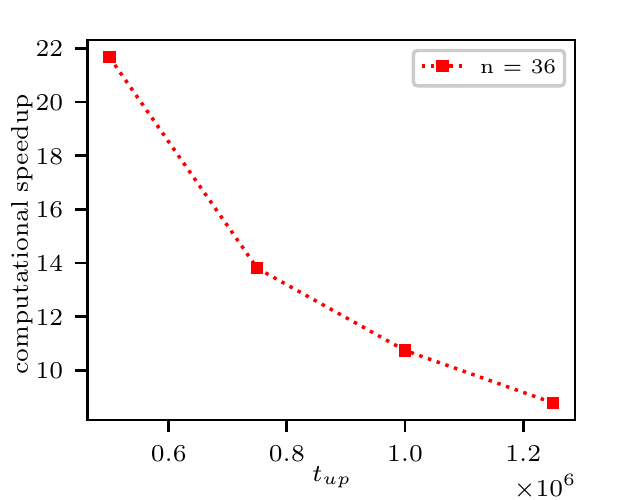}}
    \caption{Speedup of cumulants updates comparing with 
    \protect \cite{domino2017tensorsnet} recalculation, $6$ workers implementation.}
    \label{fig::cup}
\end{figure*}
In Figure \ref{fig::cup} we present the computational speedup of the update
cumulants of order $1, \ldots, d$, compared with their simple recalculation
\cite{domino2016usecum} implemented in \proglang{Julia} \cite{cum}. The main
conclusion is that the computational speedup is of about one order of magnitude.
Higher speedup is recorded for large data sets.

\section{Illustrative application}\label{s::exemplaryapp}
In this section we show a~practical application of the sliding window cumulant
calculation algorithm to analyze data that are updated in batches.
As a simple application we propose the following scenario. The initial batch of
data $\mat{X}^{(1)}$ is drawn from a multivariate Gauss 
distribution. 
Then the subsequent update
batches $\mat{X}^{(w)}_{(+)}$ are drawn from $t$-Student 
copula---a strongly non-Gaussian
distribution---having the same univariate marginal as the Gauss distribution.
The transition from a Gaussian to a non-Gaussian regime is observed using the value of 
Froebenius norm of fourth cumulant tensor.

\subsection{Cumulants based measures of data statistics}{\label{s::measureofdivgauss}}
According to the definition of high order cumulants 
\cite{kendall1946advanced,lukacs1970characteristics} they are zero only if data 
are sampled from multivariate Gaussian distribution. Hence in this case the Frobenius norm of 
high order cumulant tensor:
\begin{equation}{\label{eq::me}}
\|\tensor{C}_d\|_k = 
\sqrt[\leftroot{-1}\uproot{3}k]{\sum_{\midx{i}} |c_{\midx{i}}|^k}
\end{equation}
should be zero as well. 
Let us introduce a function 
\begin{equation}{\label{eq::mef}}
\nu_d =  \frac{\|\tensor{C}_d\|_2}{\|\tensor{C}_2\|_2^{d/2}} \text{ for } d > 2,
\end{equation}
that will be used to detect non-Gaussianity of data.
Recall that in the case of univariate random variable, for $d=3$ and $d = 4$ 
the function $\nu_d$ is equal to the modules of asymmetry and kurtosis. 
Obviously, for multivariate data, the higher values of $\nu_d$ the less likely
that data were drawn from a multivariate Gaussian distribution.

Due to the use of block structure
\cite{schatz2014exploiting,domino2017tensorsnet} the function $\nu_d$ can be
computed fast and use small amount of memory

Suppose we have the supper-symmetric cumulant tensor $\tensor{A} \in \Real^{[n,
d]}$ stored in a~block structure, \ie\ we store only one hyper-pyramidal part of
such tensor in blocks. Let $\midx{j} = (j_1, \ldots, j_d)$ be a multi-index of
block $(\tensor{A})_{\midx{j}} \in \Real^{b^d}$, without loss of generality and for
the sake of simplicity we assume that $b | n$. Then, in the block structure, we
store only blocks indexed by such $\midx{j}$ whose elements are sorted in
an increasing order.

We propose Algorithm~\ref{alg::norm} that computes a $k$-norm of given super-symmetric
tensor $\tensor{A} \in \Real ^{[n,d]}$. 
Blocks in a~block structure can be super-diagonal (super-symmetric), 
partially diagonal (partially-symmetric) or off-diagonal.
\begin{algorithm}
	\caption{Calculate $k$-norm of the tensor stored in a block structure.} 
	\label{alg::norm}
	\begin{algorithmic}[1]
		\State \textbf{Input}: $\tensor{A} \in \Real^{[n,d]}$ -- the 
		supper--symmetric 
		tensor stored in blocks, $\bar{n}$ -- number of blocks
		\State \textbf{Output:} Number -- the $k$-norm of the tensor
		\State $z\gets 0$
		\For{$j_1 \gets 1:\bar{n}, \ldots, j_{d} \gets j_{d-1}$ \text{to} 
			$\bar{n}$} 
		\State $\mathbf{j} = (j_1, \ldots, j_d)$
		\State \begin{equation*}z \gets z + \frac{d!}{\prod_l r_l!}\sum_{e 
			\in (\tensor{A})_{\mathbf{j}}} |e|^k\end{equation*}
		\Comment{$(\tensor{A})_{\mathbf{j}}$ denotes a~block indexed by 
			multi-index $\mathbf{j}$}
		\EndFor 
		\State \Return $^k\sqrt{z}$
	\end{algorithmic}
\end{algorithm}      

Let $(\tensor{A})_{\midx{j}}$ be an off-diagonal block, hence $j_1 < j_2 <
\ldots < j_d$. In order to compute the Froebenius norm its elements must be
counted $d!$ times in the sum since such block appears---up to generalized
transpositions---$d!$ times in the in the full super-symmetric tensor.
Since the Froebenius norm is an element-wise function, the order of tensor
elements is not important.

In two other cases, partially diagonal or super-diagonal blocks have repeating 
indices---theirs multi-indices $\midx{j}$ are equal to: 
\begin{equation}
\begin{split}
(j_1 < \ldots < \underbrace{j_{s_1} = \ldots = j_{s_{r_1}}}_{r_1} < \ldots <&
\\< \underbrace{j_{s_2} = \ldots = j_{s_{r_2}}}_{r_2} < \ldots &).
\end{split}
\end{equation}
Such blocks are repeated $\frac{d!}{\prod_l r_l !}$ times in the full tensor.
Note that, if $j_1 < j_2 < \ldots < j_d$ then $\prod_l r_l ! = 1$.
In the super-diagonal case \ie\ $\midx{j} = (\underbrace{j_1 = \ldots = j_d}_d)$ 
we have:
\begin{equation}
\frac{d!}{\prod_l r_l !} = \frac{d!}{d!} = 1,
\end{equation}
so the super-diagonal block is counted only once as expected.

The advantage of Algorithm~\ref{alg::norm} is that it iterates over blocks in
the block structure, what allows for efficient computation of the internal sum
elements. A~naive element-wise norm calculation approach would require $n^d$
power operations. To compute $\|\tensor{A}\|$ using Algorithm \ref{alg::norm} we
need $b^d$ power operations for each block. Taking advantage of the block
structure, required number of multiplications can be approximated by
$\frac{n^d}{d!}$. Finally, the computational complexity of Algorithm
\ref{alg::norm} is small comparing with the computational complexity of
Algorithm \ref{alg::momentsupdate}, the complexity of the procedure of cumulants
updates and computation of their norms can be approximated by
Eq.~\eqref{eq::complxn}.

\subsection{Data stream generation}{\label{s::datageneration}}
In order to illustrate the functioning of aforementioned algorithms
 we use an artificially generated stream of data.

The initial data batch $\mat{X}^{(1)} \in \Real^{t \times n}$ 
is sampled from a Gaussian multivariate distribution $\mathcal{N}(\mu, \Sigma)$, 
where ideally $\mu = \tensor{C}_1(\mat{X}^{(1)})$, $\Sigma = \tensor{C}_2(\mat{X}^{(1)})$.
The subsequent data batches $\mat{X}^{(w)}_{(+)} \in \Real^{t_{\mathrm{up}} 
\times n}$,
for $w \geq 1$ are sampled from distribution $F$, which explained 
further.
Our goal is to determine if the distribution of the updated data $\mat{X}^{(w)}
\in \Real^{t \times n}$ did not change with raising $w$,
and is still a~multivariate Gaussian.

A~naive approach would be to compute the multiple univariate statistics, such as
for example: asymmetry $\kappa_3$ and kurtosis $\kappa_4$ for each of the
marginal variables of the data stream in windows $\mat{X}^{(w)}$
\cite{gama2010knowledge}.
However such approach is oversimplified,
since from the Sklar Theorem \cite{sklar1959fonctions} one can deduce that it is
always possible to construct such non-Gaussian multivariate distribution $F$
that has all marginal distributions $(F_i)$ being univariate Gaussian. Hence despite
$\forall_{i} \ \kappa_3(F_i) = 0, \kappa_4(F_i) = 0$, $F$ is not a multivariate
Gaussian.

To generate such data in practice we can use a copula approach, see \textit{e.g.}
\cite{cherubini2004copula} for definition and formal introduction of copulas.
The probability distribution $F$ is derived from the $t$-Student copula parametrized by 
$\Sigma$ and $\nu$ as defined in \cite{cherubini2004copula}.
In our case we set the parameter to $\nu = 10$ degrees of freedom
and the marginals equal to those of $\mat{X}^{(1)}$.

\begin{figure}[t!]
    \centering
        \includegraphics{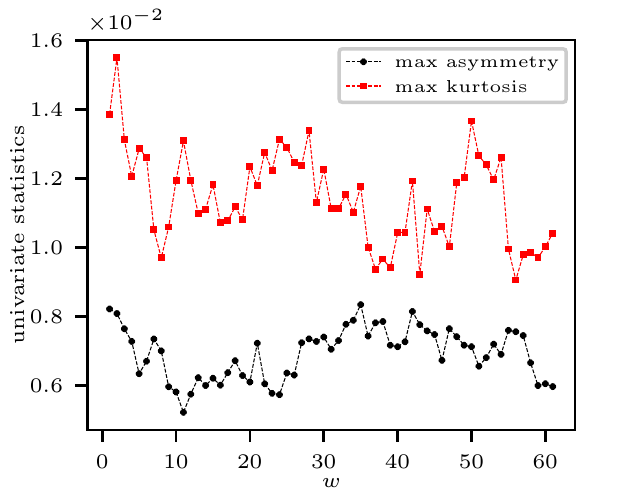}
    \caption{Maximums of absolute values of univariate asymmetries and kurtosises for 
    $\mat{X}^{(w)}$ with number of marginal values $n = 60$, $t = 10^6$ data 
    samples, $t_{\text{up}} = 2.5 \cdot 10^4$, $w_{\max} = 61$.}
    \label{fig::1ds}
\end{figure}

In order to visualize statistics of the generated data we 
calculate the maximums over marginals of absolute values of univariate asymmetries
and kurtosises for $\mat{X}^{(w)}$. The results are presented in Fig.~\ref{fig::1ds},
as discussed before neither univariate asymmetry nor kurtosis is 
significantly affected by the update.

\subsection{Stream statistics analysis}
\begin{figure}[t!]
    \centering
    \subfigure[$\|C_2\|$,\label{fig::nc1}]%
    {\includegraphics[width=0.4\textwidth]{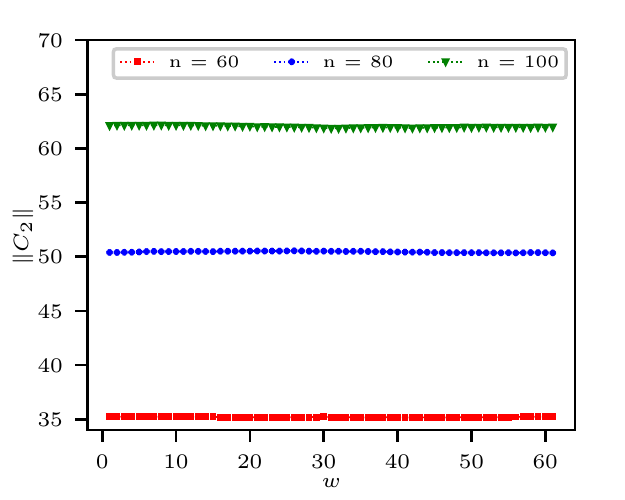}}
    \subfigure[$\nu_3(\mat{X}^{(w)}) \ \nu_4(\mat{X}^{(w)}) \ n = 
    100,$\label{fig::nc2}]%
    {\includegraphics[width=0.4\textwidth]{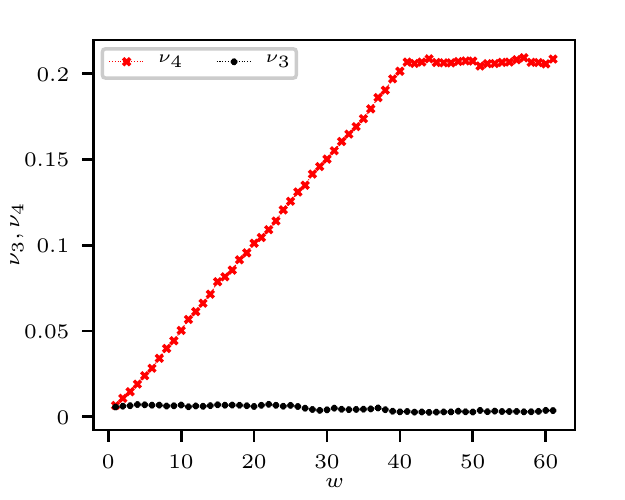}}
    \caption{Cumulants based statistical measures for data $\mat{X}^{(w)} \in 
    \Real^{t \times n}$, with window width $t = 10^6$, update width $t_{\text{up}} = 2.5 \cdot 10^4$, and number of windows
    $w_{\max} = 61$.
    The initial data are drawn from a~Gaussian distribution, then the 
    subsequent updates are drawn from a non-Gaussian one. We can observe that, 
    in this case, $\nu_4$ is a good estimator of non-Gaussianity in contrast to 
    function $\nu_3$. The value of $\nu_4$ raises with each window up to the 
    saturation point at $w = 41$.
    }
    \label{fig::nc}
\end{figure}

{\label{s::streamanalysis}}
In order to detect the change in the probability distribution we calculate the
following values of cumulants based measures in function of $w$. Those measures
are $\|\tensor{C}_2(\mat{X}^{(w)})\|$, $\nu_3(\mat{X}^{(w)})$ and
$\nu_4(\mat{X}^{(w)})$, see Eq.~\eqref{eq::mef}. The obtained results are
gathered in Fig.~\ref{fig::nc}. Analyzing the panel~\ref{fig::nc1} one can see
that the norm of the covariance matrix is not significantly affected by the
updates, this is due to the particular choice of the t-Student copula parameters
\cite{cherubini2004copula} used to generate the updates.

Further, as presented in panel \ref{fig::nc2} the $\nu_3(\mat{X}^{(w)})$ is also 
unaffected by updates, because t-Student
copula is symmetric \cite{cherubini2004copula} in such a way that, given 
symmetric marginals, its high order odd cumulants are zero. However given 
t-Student copula this is not the
case for even cumulants, \eg $\nu_4(\mat{X}^{(w)})$ is strongly affected by updates
and in this case can by used to distinguish between underling distributions from which data are drawn. 
The values of $\nu_4(\mat{X}^{(w)})$ raise with raising window number $w$, up 
to $w = 41$, since for $w > 41$ there is no original data from multivariate Gaussian distribution
left in $\mat{X}^{(w)}$.

The normalization factor in the denominator of $\nu_4$ assures that 
the function behaves similarly for different number of marginal variables $n$. 
This behavior depends from the particular choice of the $t$-Student copula used in updates, 
however in general the choice of the particular measure $\nu_d$ should depend on the expected 
statistical model of a~data stream.

Let us discuss the approximation error of $\nu_{4}(\mathbf{X})$. We assume that 
estimation error of $d$\textsuperscript{th} cumulant elements comes mainly from 
estimation error of corresponding $d$\textsuperscript{th} moment element. In an 
super-diagonal case, we can refer directly to 
Appendix~$A$ in \cite{domino2017tensorsnet} and recall that the standard error 
of the estimation of $d$\textsuperscript{th} univariate moment $m_d$ is limited 
by $\sqrt{\frac{m_{2d}}{t}}$, in our case it is limited by
$\sqrt{\frac{7!!}{10^6}} \approx 10^{-2}$. In a case of 
off-diagonal elements of $\mathcal{C}_4$, as mentioned in aforementioned Appendix~$A$, 
estimation error is limited by 
the product of lower order moments which generally should by limited by 
$m_{2d}$, since moments' values rise rapidly with $d$. Finally, while 
computing $||\mathcal{C}_4||$, see~\eqref{eq::me}, we sum up squares of its 
elements,  hence their individual errors should cancel out to some extend. 
However dependency between those elements is complex and a~standard 
error calculus would be complicated. 
Hence we have performed $100$ numerical experiment, and computed for $n = 100$ 
and $t = 10^6$ $\nu_4$ from generated data. We obtained the following results 
summarized by the triplet of values: $5$\textsuperscript{th} quantile, 
median and $95$\textsuperscript{th} quantile of $\nu_4$ values. 
At $w = 1$---Gaussian multivariate distribution---we have obtained 
$(0.004, 0.006, 0.011)$, while at $w \geq 41$---t-Student copula 
with Gaussian marginals---we have obtained $(0.199, 0.209, 0.220)$. 
In the second case the error is higher since the 
t-Student copula introduces high order dependencies between data and elements 
of $\mathcal{C}_4$. Concluding the estimation error is small 
in comparison with $\nu_4$ values.

\subsection{Data frequency analysis}\label{s::datafreqanalysis}
In order to estimate the maximal frequency of a data stream that can be analyzed
on-line using Algorithm~\ref{alg::mainloop} we perform the following experiment
using the same hardware discussed in Section~\ref{s::tests}. 
We fix the number of samples in a observation window $t$ and vary the
number of marginals $n$ and number of samples in a~batch $t_\mathrm{up}$.
After the line~\ref{ln::emitcumulants} of Algorithm~\ref{alg::mainloop} is 
executed values of 
$\|\tensor{C}_1(\mat{X}^{(w)}) \|$, 
$\| \tensor{C}_2(\mat{X}^{(w)}) \|$, 
$\nu_3(\mat{X}^{(w)})$ and $\nu_4(\mat{X}^{(w)})$ are calculated.

\begin{figure}[h]
    \centering
    \includegraphics[width=0.4\textwidth]{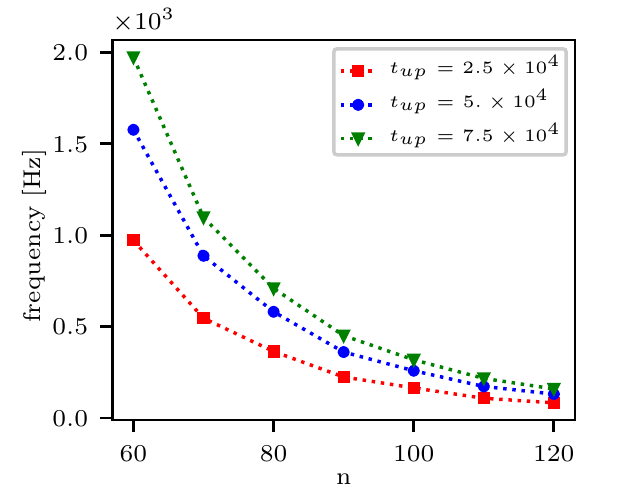}
    \caption{\label{fig::fa} Maximum frequency of on-line data analyzed using the cumulants 
    updates scheme, $d = 4$, $b = 4$, multiprocessing computation on $6$ 
    workers.}
\end{figure}

In Figure \ref{fig::fa} we present the maximal frequency of data analyzed
on-line using the proposed scheme. In presented example we compute and update
cumulants of order $1, \ldots, 4$ and use $\|\tensor{C}_1(\mat{X}^{(w)}) \|$, $\| \tensor{C}_2(\mat{X}^{(w)}) \|$,
$\nu_3(\mat{X}^{(w)})$ and $\nu_4(\mat{X}^{(w)})$ to extract statistical features. 

Consider that the Algorithm~\ref{alg::symoutprod} for cumulants' updates, is
independent of $t_{\mathrm{up}}$ and therefore has constant execution time. 
Therefore  one can increase maximal data frequency at the expanse of the method
sensitivity by increasing $t_{\mathrm{up}}$.

\section{Conclusions}
In this paper we have introduced a
sliding window cumulant calculation algorithm
for processing on-line high frequency multivariate data. 
For computer hardware
described in Section \ref{s::implementation} we have obtained maximum data processing 
frequency of $150$--$2000$ Hz depending on a number of marginal variables. 
We have presented an illustratory application of our algorithm by employing an example of
Gaussian distributed data updated by data generated using t-Student copula. We 
have shown that our algorithm can be used successfully to determine if on-line 
updates break the Gaussian distribution. 

We believe that presented algorithm can find many new applications for example
in on-line signal filtering or classification of data streams. The algorithm can be
combined with many different methods of the cumulant-based statistical features
extractions, such as the Independent Component Analysis
(ICA)~\cite{blaschke2004cubica,virta2015joint} or based on tensor eigenvalues
\cite{qi2005eigenvalues}.

\section*{Acknowledgments}
The research was partially financed by the National Science Centre,
Poland---project number 2014/15/B/ST6/05204. The authors would like to thank
Przemys\l{}aw G\l{}omb for reading the manuscript and providing valuable
discussion. 
\bibliographystyle{ieeetr}
\bibliography{cumulants_up}

\end{document}